# High Performance 5G FR-2 Millimeter-Wave Antenna Array for Point-to-Point and Point-to-Multipoint Operation: Design and OTA Measurements Using a Compact Antenna Test Range

Abdul Jabbar[1*], Jalil Ur-Rehman Kazim[2], Mahmoud A. Shawky[2], Muhammad Ali Imran[2], Qammer Abbasi[2], Muhammad Usman[1], and Masood Ur-Rehman[2]

[1]Department of Electrical and Electronic Engineering, Glasgow Caledonian University, UK

[2]James Watt School of Engineering, University of Glasgow, UK

*Corresponding Author: abduljabbar@ieee.org

*Abstract*—This paper presents the design and comprehensive measurements of a high-performance 8-element linear array and a compact high-gain 32-element planar antenna array covering the n257 (26.5–29.5 GHz) FR-2 millimeter-wave (mmWave) band. First, an 8-element series-fed linear array is designed with a fan-shaped pattern for 5G point-to-multipoint connectivity. Then a 4-way corporate-series feed network is designed for a high-gain 32-element compact and directive array for point-to-point mmWave connectivity. Comprehensive over-the-air (OTA) measurements are conducted using a state-of-the-art compact antenna test range (CATR) system, enabling precise characterization of radiation patterns across a 180° span in the azimuth and elevation planes. The planar array achieves a peak measured gain of 18.45 dBi at 28.5 GHz, with half-power beamwidths ranging from 11°–13° (wide axis) and 23°–27° (narrow axis) across the band of interest. The sidelobe levels are below -10 dB in the desired band of interest. The measured results match well with the simulation results. The designed antenna array is applicable to various emerging 5G and beyond mmWave applications such as high data rate mmWave wireless backhaul, mmWave near-field focusing, high-resolution indoor radar systems, 28 GHz Local Multipoint Distribution Service (LMDS) as well as the characterization of mmWave path loss and channel sounding in diverse indoor environments.

## I. Introduction

The development of fifth-generation (5G) and beyond communication technology has marked a distinguished era in wireless communication, characterized by significantly enhanced connectivity, reduced latency, and substantially increased data rate [1], [2]. The frequency range 2 (FR-2) spans the millimeter-wave (mmWave) band under 5G New Radio (NR) and aims to enable ultra-low-latency and wideband services, opening up a whole new era of applications and services [3], [4]. This is due to its potential to facilitate beamforming gain and gigabit-per-second data rates, necessary for bandwidth-hungry applications such as high-definition video streaming, virtual reality (VR), autonomous vehicles, short-range high-resolution radars, and indoor communication and sensing for smart factories. Moreover, while fiber optic data transfer links can provide multi-gigabit-per-second data rates, their cost and deployment are often prohibitive in many applications. Wireless links, on the contrary, can provide a cost-effective fiber alternative to interconnect the outlining areas beyond the reach of the fiber rollout. Fixed wireless access (FWA) mmWave backhaul links at mmWave is an interesting solution for this [5]. Similarly, Local Multipoint Distribution Service (LMDS) is a terrestrial point-to-multipoint (P2MP) radio service providing wireless broadband access to fixed networks, where 28 GHz antennas and radio units find huge potential [6]. This necessitates highly directional antennas and sometimes steerable antenna beams to compensate for the high propagation loss. The mmWave transmission is envisaged as a key enabling technology for future multi-cell large-scale antenna systems where mmWave antennas with small sizes allow for dense antenna packing in small areas for high beamforming gains [7]. The high frequency translates into shorter wavelengths, which facilitates the design of compact, high-gain antenna arrays. Nevertheless, it also presents challenges such as higher propagation losses and susceptibility to atmospheric absorption, necessitating efficient and directional antenna designs to ensure robust network performance [8], [9]. Thus mmWave applications require directional antennas at edge node/base stations as well as at customer premises equipment (CPE) terminals (with specified radiation patterns), as depicted in Fig. 1.

Various types of directional antenna designs are reported at mmWave bands such as planar substrate integrated waveguide (SIW) based antennas [10], 3D printed antennas [11], and other versatile printed circuited board (PCB) based antennas [12]–[23] Amongst these, PCB-based series-fed microstrip antenna technology is well-suited for many of the 5G and beyond application scenarios for its low profile, lightweight, ease of fabrication, and seamless integration with radio frontends [24]–[27]. When configured as an array, microstrip antennas can achieve higher gain and directivity, crucial for overcoming the higher path loss at mmWave bands and enhancing signal-to-noise ratio (SNR). Owing to the smaller wavelength at mmWave bands, the compact antenna arrays offer high directionality and beamforming gain, essential for directing the antenna's radiation pattern toward the desired user, thus improving network performance and connectivity.

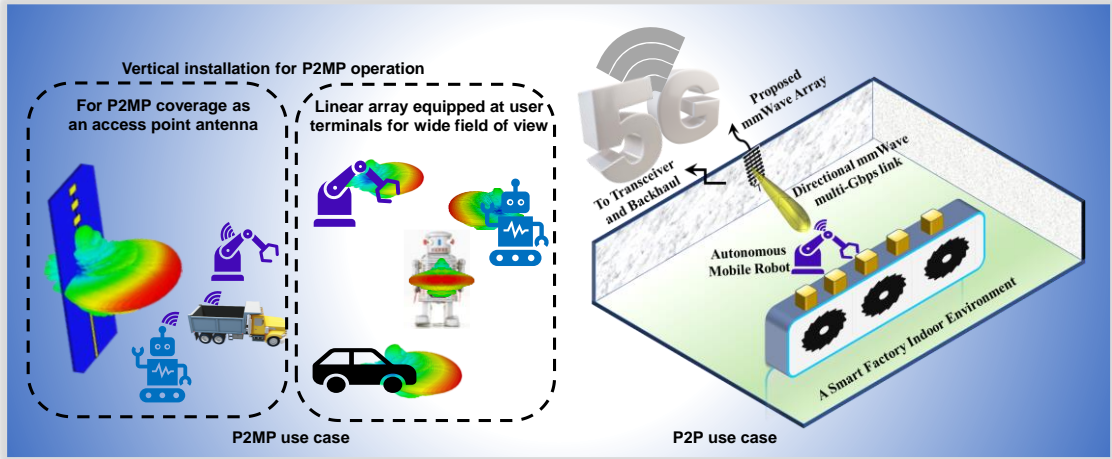

Fig. 1. A conceptual illustration of P2P and P2MP mmWave communication operation.

Another critical aspect of mmWave antennas and devices is their measurements. Given the stringent requirements for high-resolution measurements and the sensitivity of mmWave frequencies to environmental factors, a compact antenna test range (CATR) measurement system offers a controlled and efficient solution for far-field testing in a compact space. This method ensures precise measurement of critical parameters such as gain, radiation patterns, and beam-steering capabilities, which are vital for validating the design and functionality of mmWave antenna arrays. When compared with the direct far-field method, the CATR can significantly reduce the system's path loss [28], making it the most fitting and appealing test environment for mmWave applications such as beamforming antenna arrays and devices [29]–[33], 5G base station evaluations [34] as well as mmWave channel-sounding [35].

In this paper, we address these challenges and present the design, implementation, and detailed measurement of a compact and high-directivity 28 GHz FR-2 planar microstrip antenna array, featuring a connectorized solution. The proposed antenna simplifies the design process by avoiding any of the complex features such as inset cuts, slots, vias, and parasitic patches, resulting in ultra-low-cost fabrication and seamless compatibility with standard RF equipment without compromising performance. The design is optimized for enhanced gain, matched bandwidth, compact size, and high radiation efficiency, aiming to boost 5G network performance through reliable and consistent operation. The 8-element linear array produces a fan-beam radiation pattern, making it ideal for P2MP mmWave applications. The 32-element planar array provides a narrow beam with high gain, covering the n257 5G band (26.5–29.5 GHz), making it suitable for point-to-point (P2P) connectivity. The array employs a solderless edge-fed RF feed mechanism, which enhances reusability and offers a simple, cost-effective feeding solution. This approach minimizes losses typically associated with soldered connections at mmWave frequencies and reduces complexity compared to other feeding techniques. Additionally, the edge-launched microstrip feed ensures seamless integration with mmWave connectorized systems and hence simplifies device characterization, supporting robust and efficient deployment in advanced mmWave communication systems.

In addition, the proposed antenna serves as a viable and cost-effective alternative to expensive mmWave horn antennas, offering comparable high-gain performance and directional radiation characteristics at a fraction of the cost. Unlike horn antennas, which are bulky, costly, and often challenging to integrate into compact systems, the presented planar microstrip array is lightweight, low-profile, and easy to fabricate, making it suitable for mass production and practical deployment in mmWave licensed 5G networks.

Further, we elucidate the comprehensive radiation pattern and gain measurements of the antenna array using Rohde & Schwarz (R&S) ATS800B® CATR setup. Furthermore, the detailed measurement methodology, covering gain, radiation patterns, and beamwidth, demonstrates the reliability of the design and offers valuable insights into OTA mmWave antenna evaluation processes, facilitating robust antenna performance validation for 5G and beyond technologies. This is a significant reduction in distance compared to the standard antenna test configuration, which typically requires a minimum distance of $2D^2/\lambda$ to achieve the same size test region. These benefits encompass the ability to directly evaluate far-field attributes of diverse antenna array configurations across a broad frequency spectrum, within a confined testing space. The proposed antenna design and measurement procedure offer valuable insights for practical OTA mmWave antenna array evaluations, encompassing radiation pattern analysis and realized gain measurements, catering to a wide range of mmWave applications.

## II. Design and Characterization of Millimeter-Wave Antenna Array

The design perspective of the proposed antenna array is shown in Fig. 2. The antenna is simulated in full-wave CST electromagnetic solver using RO3003 substrate with a thickness of 0.25 mm, dielectric constant of 3, loss tangent of 0.001, and copper metal thickness ($m_t$) of 18 μm. Initially, a patch antenna element is designed on the resonance band of 28 GHz. Although a single patch antenna is inherently narrowband, the collective response of multiple resonances of various such series-fed patch elements superimposes to form a wideband impedance bandwidth [36]. Then further patch elements were added to the first element in a series-fed topology and reflection bandwidth was analyzed. The design evolution is illustrated in Fig. 3(a) along with results of -10 dB impedance bandwidth. In this approach, the widths and lengths are analytically calculated and tuned to achieve a linear array of patches that radiate with the desired amplitudes and phases around the 28 GHz band. A small tapering is applied to the width of the first patch element to further fine-tune the -10 dB impedance bandwidth. The patch elements are spaced at a distance of approximately $\lambda_0/2$ (where $\lambda_0$ is the free space wavelength at 28 GHz) to ensure optimal performance and minimize mutual coupling. The design initially ran through an extensive analytical and simulation process for its critical parameters such as the length and width of the patch antennas, the main feed width, and the inter-element length.

For the design evolution steps of Fig. 2(a), the width ($W$) and length ($L$) of the microstrip patch element at 28 GHz were calculated using standard patch antenna equations as [37]:

$$W = \frac{c}{2f_r}\sqrt{\frac{2}{\varepsilon_r+1}} \tag{1}$$

$$L = \frac{c}{2f_r\sqrt{\varepsilon_{eff}}} - 2\Delta L \tag{2}$$

Here $f_r$ is the desired operational frequency, $c$ is the speed of light, $\varepsilon_r$ is the relative permittivity of the substrate, $\varepsilon_{eff}$ is the effective relative permittivity of the substrate, $h$ is the thickness of the substrate, and $\Delta L$ is the extended length of the patch that accounts for the fringing effect and depends on the dielectric constant, width, and height of the substrate. The effective permittivity and the extended length are calculated as:

$$\varepsilon_{eff} = \frac{\varepsilon_r+1}{2} + \frac{\varepsilon_r-1}{2}[1 + 12\frac{h}{W}]^{-\frac{1}{2}} \tag{3}$$

$$\Delta L = 0.412h\frac{(\varepsilon_{eff}+0.3)(\frac{W}{h}+0.264)}{(\varepsilon_{eff}-0.258)(\frac{W}{h}+0.8)} \tag{4}$$

At mmWave frequencies, the edge-launched feed offers easy integration and facilitates straightforward antenna measurements. The width of the main feed line was matched to 50 Ω ($Z_0$) calculated using (5) and then further fine-tuned to the optimized value of 0.623 mm.

$$w_f = \frac{7.48 \times H}{e^{Z_0\frac{\sqrt{(\varepsilon_r+1.41)}}{87}}} - (1.25 \times m_t) \tag{5}$$

An extensive parametric sweep was applied during the design evolution to the width of patch elements, as well as to the width and length of the series-connected feed lines. The lengths and widths of the whole linear array were collectively fine-tuned to achieve a -10 dB impedance bandwidth of at least 3 GHz around 28 GHz, thereby covering the 26.5–29.5 GHz band with the desired radiation pattern and gain characteristics, ensuring robust performance for mmWave applications. It is worth mentioning here that the inter-element gap in this series-fed linear array has a significant impact on the input impedance and overall antenna performance, as illustrated in Fig. 2(b). During the design process, we thoroughly investigated the effect of varying the inter-element gap (by keeping the width of patches constant) on key performance parameters such as impedance matching, gain, and efficiency. The gap defines the electrical length of the whole linear array and governs the progressive phase shifts between array elements. Hence the phase constant can change accordingly which has an impact on the input impedance as well as the radiation pattern (particularly beam-squint) of the array. The optimal gap value was fine-tuned to be 2.82 mm, which provided the finest trade-off between these factors in this design. The dimensions of the fine-tuned linear array are shown in Fig. 2(c).

To achieve a compact planar array form factor, a 4-channel equal power divider is designed as shown in Fig. 2(d). The 4-way feed network is designed with appropriate impedance matching to ensure efficient power distribution to each element. The edges of the power divider are chamfered to reduce the spurious emissions through the corner. The interelement spacing ($g$) between the arms of the parallel feed network (and thus between the center of antenna elements) is set as 5.34 mm ($\lambda_0/2$) to minimize the mutual coupling as much as possible (here $\lambda_0$ is the free space wavelength at 28 GHz). Then, for high gain and directional beams, each 8-element linear array is utilized as a sub-array with the output ports of the power divider to create a 32-element (4 × 8) compact corporate-series planar array as shown in Fig. 2(e). The dimensions of the proposed antenna array are presented in Table 1.

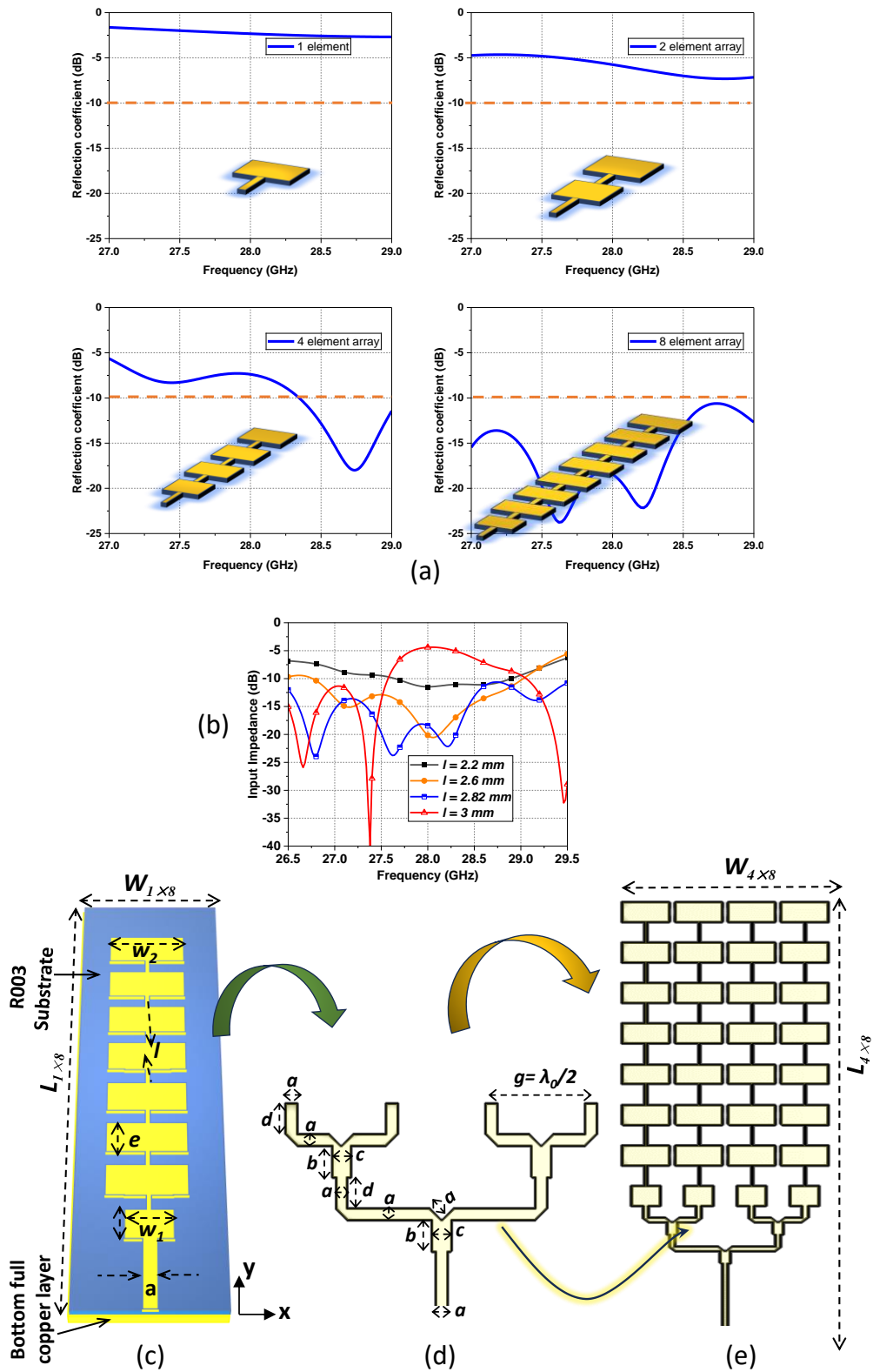

Fig. 2. (a) Design evolution steps of the proposed array with the effect of increasing the number of patch elements on reflection coefficients of the linear array. (b) Effect of the inter-element gap on the input impedance of the linear array. (c) Schematic of the optimized proposed 8-element linear antenna array. (d) 4-way mmWave equal power divider. (e) Schematic design of the proposed 32-element planar array.

TABLE 1. Geometric parameters (in millimeters) of the proposed antenna array design.

| *a* | *b* | *c* | *d* |
|---|---|---|---|
| *0.62* | *1.69* | *1.04* | *1.67* |
| *e* | *l* | *$w_1$* | *$w_2$* |
| *2.80* | *2.82* | *3* | *4.8* |
| *$W_{1\times8}$* | *$L_{1\times8}$* | *$W_{4\times8}$* | *$L_{4\times8}$* |
| *13* | *52* | *23* | *60* |

### III. FULL-WAVE SIMULATION RESULTS AND ANALYSIS

As shown in Fig. 3(a), the -10 dB impedance bandwidth of both linear and planar arrays covers 26.5–29.5 GHz (n257 FR-2 mmWave band). The simulated peak realized gain of linear is 14.45 dBi at 29 GHz, whereas for the planar array is 18.5 dBi at 28.5 GHz respectively as shown in Fig. 3(b). The variation in gain is within 1.5 dB for the planar array in the band of interest. The radiation efficiency is within 83% to 85.7% for linear array, and 80.3% to 85.22% for the planar array. The total efficiency is above 81.47% for the linear array and above 80% for the planar array within 26.5 GHz to 28 GHz.

It is important to mention that the consistency in the magnitude and phase of the transmission coefficients of a power divider is crucial for the reliable transmission of data packets for accurate delivery of signals to the nodes of the 5G access network. Note that for a 4-port power divider, ideally, 1/4th of power should reach each output port, which is equivalent to -6 dB. The insertion loss of the designed power divider (i.e., the magnitude of transmission coefficients S21, S31, S41, and S51) is within −6.68 to −6.56 dB, whereas the reflection coefficient (|S11|) is maintained well below -17 dB in the whole band of interest. The magnitude of reflection (S11) and transmission coefficients is shown in Fig. 4(a), while the phase response of the transmission coefficients is shown in Fig. 4(b).

With the reference antenna orientation as of Fig. 5(a), the linear array provides a fan-shaped radiation pattern with wider half-power beamwidth (HPBW) ranging between 70° to 85° in the azimuth (x-z) plane and narrow HPBW in the elevation (y-z) plane ranging between 11.8° to 16°. The HPBW of the planar array in the x-z plane varies between 23° to 28°, whereas in the y-z plane, it varies between 10° to 13.2°, within the band of interest. A more directive beam is obtained from planar array due to planar geometry and more number of radiating elements. The sidelobe levels are well below -10 dB. The cross-polarization (X-pol) levels are below -20 dB for both arrays. The 3-D radiation patterns for both arrays are presented in Fig. 5 (a). A 3-D waterfall plot of the planar array in the elevation plane is illustrated in Fig. 5 (b). Note that the fan-shaped radiation pattern is desirable for multi-point coverage such as in Internet of Things (IoT) and P2MP 5G coverage, whereas a narrow directive beam is desirable for P2P high data rate mmWave links.

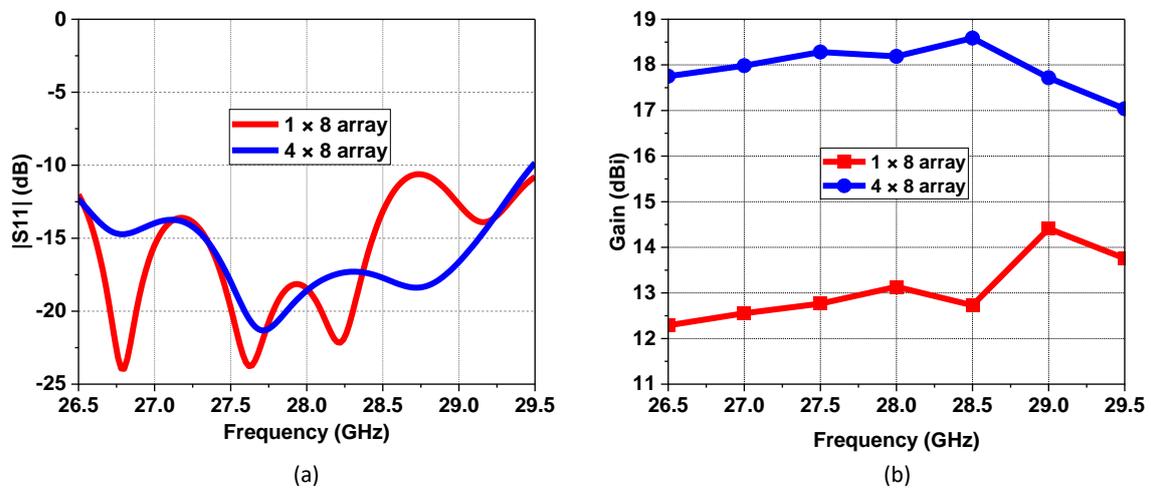

Fig. 3. (a) Simulated reflection coefficient of linear and planar arrays. (b) Realized gain of the designed arrays.

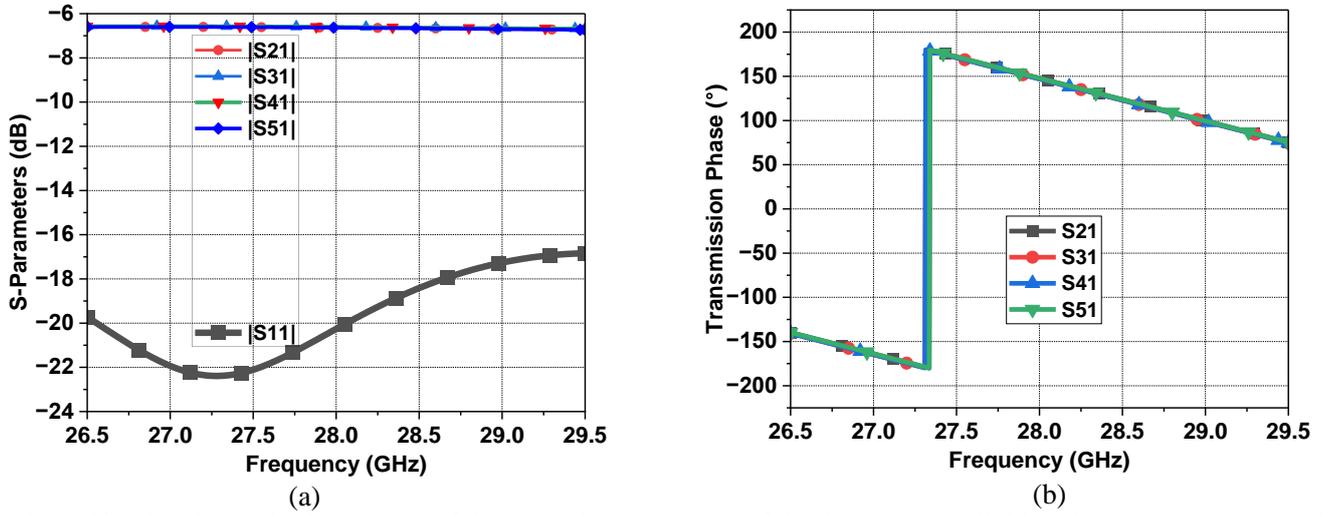

Fig. 4. (a) Simulated magnitude response of the scattering parameters of the 4-way power divider. (b) Phase response of the transmission coefficients of the power divider.

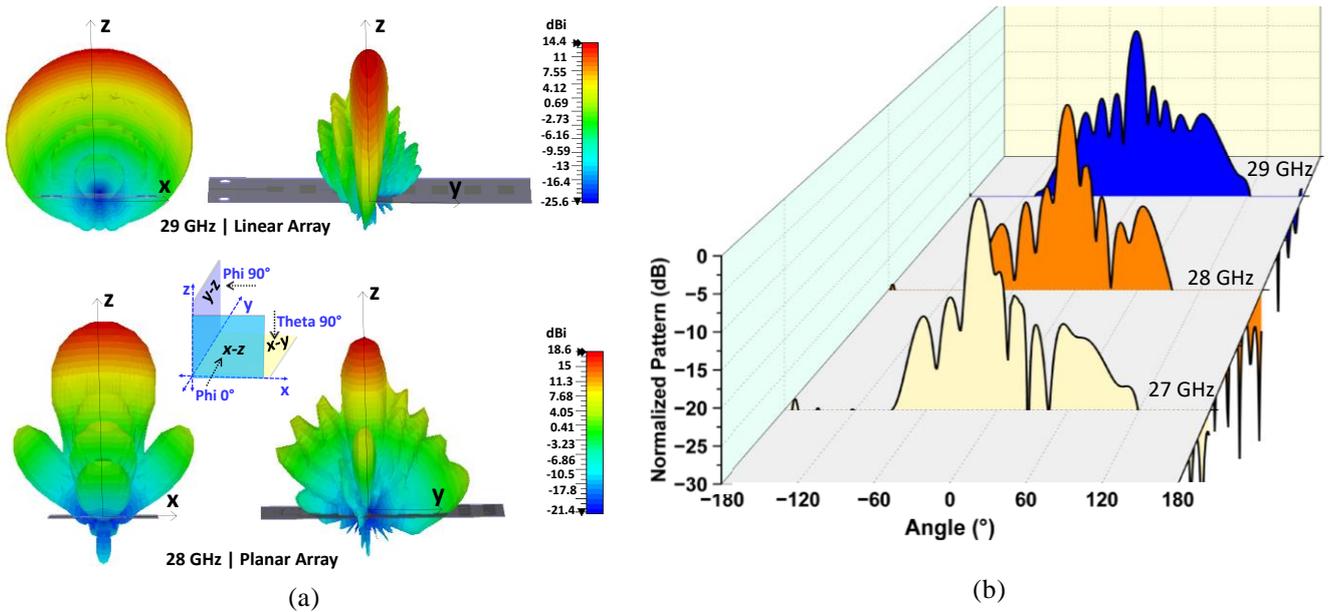

Fig. 5. (a) 3-D radiation patterns of the proposed linear and planar antenna arrays. (b) 3-D waterfall graphs of the planar array in elevation plane.

## IV. Prototype Fabrication and OTA Radiation Pattern and Gain Measurements on CATR

The prototypes of both antenna arrays were fabricated using an LPKF milling machine. The measurement setup for the reflection coefficient of both arrays is shown in Fig 6. The measured results match well with the simulated response, as shown in Fig. 7. Some discrepancies are due to fabrication tolerance, surface roughness, and probable variation in the actual dielectric constant at the mmWave band.

The contemporary 5G and beyond applications necessitate a balance of performance, simplicity of fabrication, and integration capabilities, for which microstrip antenna arrays are a prominent candidate. In particular, 5G new radio in mmWave bands necessitates the radiation performance assessments through over-the-air (OTA) testing, rather than conventional conducted methods [38], [39]. This is because, at mmWave frequencies, the antenna size predominantly influences the far-field distance. Consequently, for larger mmWave antenna arrays, the far-field distance increases in proportion to the square of the antenna size along with a significant increase in path loss [33]. This renders direct far-field measurements impractical for larger antenna arrays. Furthermore, mmWave antennas are usually integrated with RF frontends, and thus lack of accessibility to the individual antenna connectors renders conventional testing methods impractical [33], [34]. To put this into physical context, consider a mmWave antenna array with a maximum dimension of 15 cm. At 28 GHz, a minimum separation of 4.2 meters would be necessary for its far-field testing. However, this extended measurement distance would result in a substantial free-space path loss of about 73.9 dB, posing challenges for a system with a restricted dynamic range. Hence, due to the deterrent effects of conventional direct far-field measurement methods for large mmWave arrays and devices, assessing the performance of mmWave devices for 5G and beyond necessitates an alternative measurement approach, such as OTA indirect test methods [28]. CATR is one of the 3GPP-approved [40] indirect far-field test methods, as demonstrated in Fig. 8.

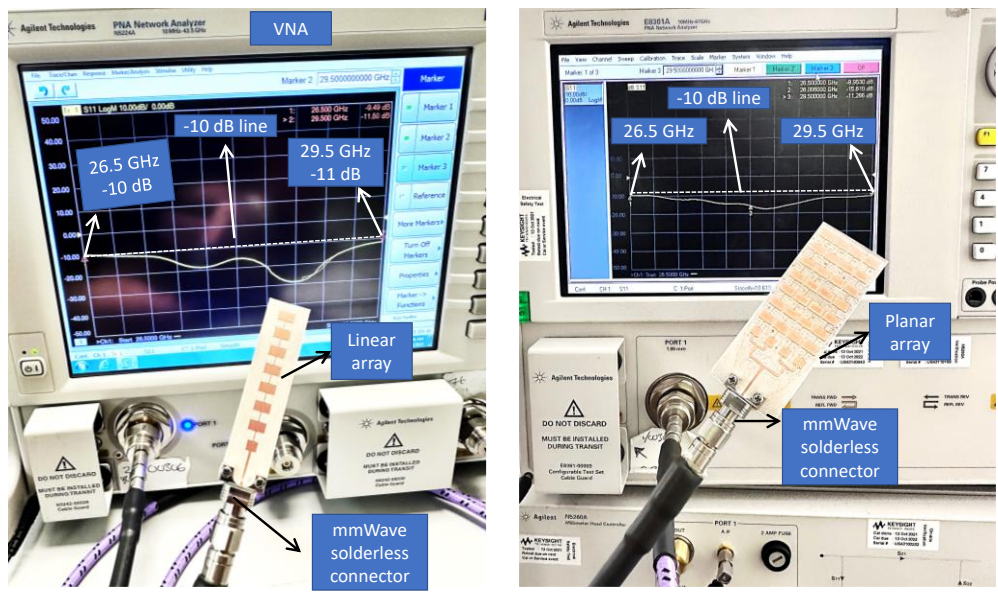

Fig. 6. Measurement of reflection coefficient (|S11|) of the fabricated prototypes of linear and planar arrays on VNA.

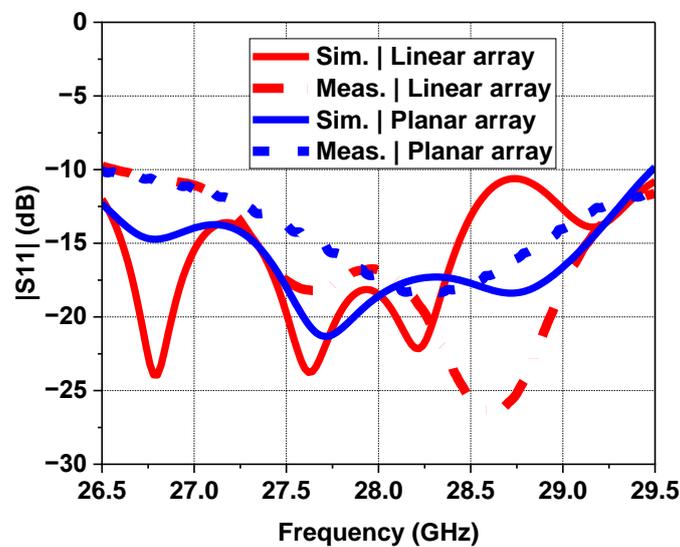

Fig. 7. Measured reflection coefficient of the linear and planar arrays.

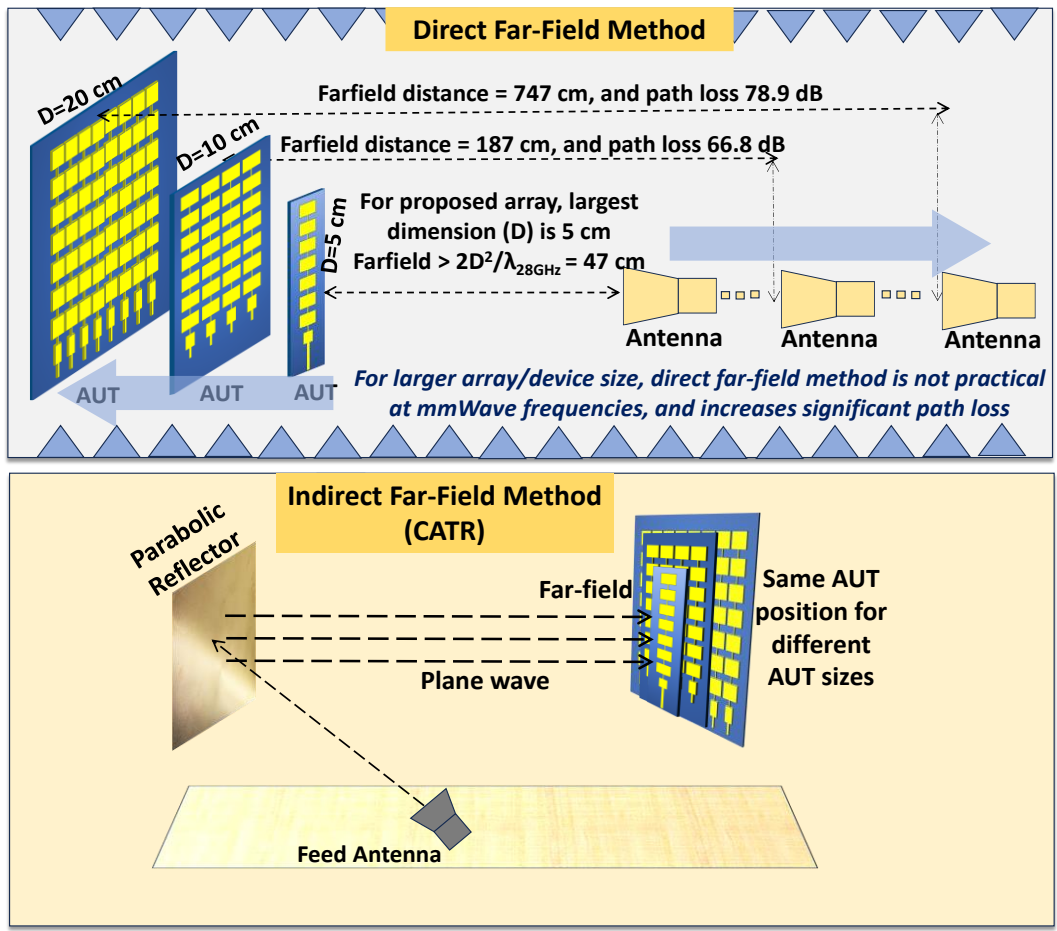

Fig. 8. A conceptual depiction of direct far-field and indirect far-field antenna measurement methods and the advantages of CATR.

In this work, we utilized the R&S®ATS800B CATR benchtop setup, as shown in Fig. 9. The CATR leverages the optical collimation attributes of a parabolic reflector to transform the spherical wavefront originating from the feed antenna into a planar wave at a very short distance towards the Antenna Under Test (AUT). This process effectively achieves a far-field region at a relatively shorter distance. The limited spatial region where the collimated wave directed by the reflector generates uniform plane waves is known as the quiet zone (QZ). This zone is situated much closer to the reflector than the far-field distance of the feed antenna alone. The measurement setup includes an ultra-wideband dual polarized transmit (Tx) Vivaldi antenna which emits a spherical wave positioned at the focal point of the parabolic reflector. This spherical wave is then converted into a plane wave by the parabolic reflector in a QZ area of 20 cm. The AUT is affixed to a rotator whose angular rotation is controlled by a PI controller motor, enabling a complete 360° rotation for angular adjustment of the AUT. A 2.4 mm standard Agilent N5224A vector network analyzer (VNA) was used to capture the S21 magnitude. Other equipment includes 2.4 mm standard connectors, right-angled adapters, and cables. This measurement system is portable and can be placed at any convenient place in an open space. As the main source of reflections in mmWave bands is mainly from the ground, therefore the ground and back side of AUT are covered with absorber cones, as can be noticed in Fig. 9. Because of high path loss and the directional nature of mmWave array antennas, sidewall reflections are not significant.

To control the VNA and the angular rotation of the positioner, we designed a custom-made LabVIEW program to carry out instrument control using SCPI commands (Standard Commands for Programmable Instruments) communicated through the VISA (Virtual Instrument Software Architecture) controller protocol. A standard NI GPIB-USB cable interface was used to connect the VNA to the PC, while the PI motor controller was connected to the PC using a USB interface. A continuous wave (CW) signal was generated using the LabVIEW program at each desired operating frequency and the radiation pattern measurement (i.e., |S21|) was automatically recorded at precise angular positions over 180° with a 1° step size (from -90° to +90°). Although full 360° measurements can be done, however, the back side of the designed antenna is fully grounded (high front-to-back ratio), therefore the back half of the array was omitted in the measurements and is not significant.

The radiation pattern was measured at three different frequencies, i.e., 27, 28, and 29 GHz. The measurement setup is shown in Fig. 10. The AUT was first mounted on the positioner (connected to port-2 of the VNA) and S21 was analyzed for a CW signal of 27 GHz. Since we utilized only one polarization of the Tx antenna at a time, it is important to ensure that the value of S21 is optimized for maximum performance before starting the measurement. This can be easily verified by testing each cable of the dual-polarized Tx antenna individually to determine which configuration yields the highest S21 value. This configuration would mean that the polarization of the Tx antenna and AUT match (i.e., co-pol case). The X-pol case was obtained by swapping the cable of the Tx antenna. Otherwise, it can also be obtained by simply rotating either of the Tx or Rx antenna by 90° and performing the same set of measurements.

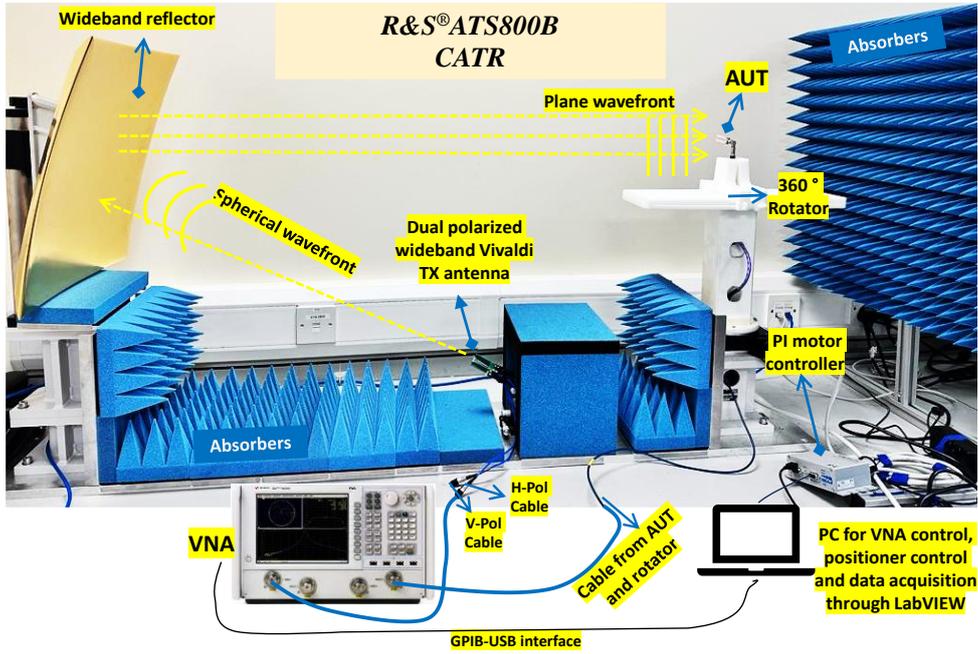

Fig. 9. OTA radiation pattern and gain measurement setup using R&S ATS800B CATR system.

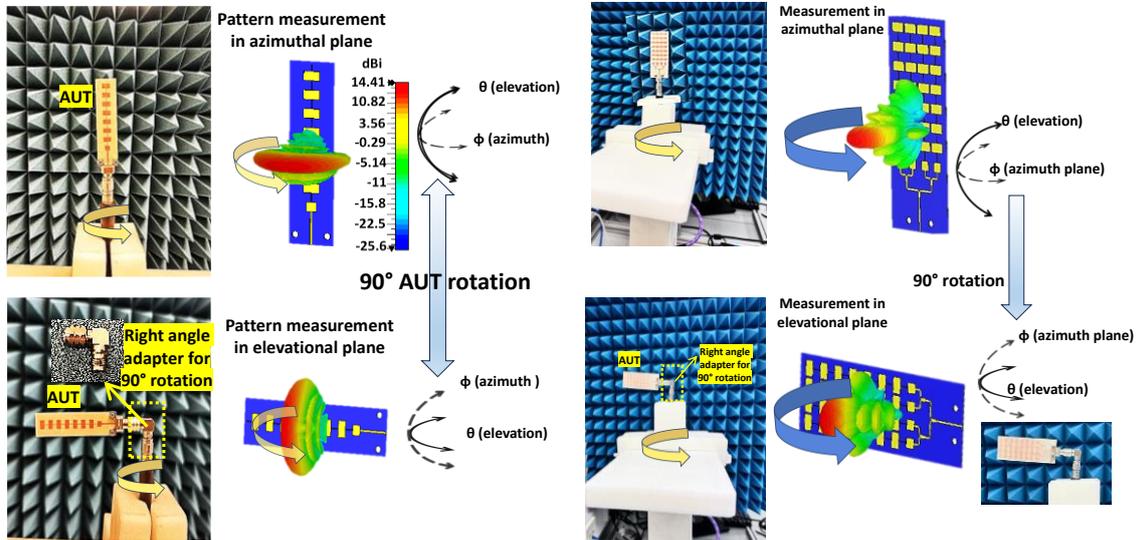

Fig. 10. Demonstration of antenna orientations for measurement of both azimuthal and elevational patterns.

The dynamic range of the CATR system was around 35 dB, which was sufficient to measure the radiation patterns and null depths. First, AUT was mounted vertically on the rotator. As shown in Fig. 10, this orientation of the antenna provided wider HPBW and thus it can be referred to as x-z cut (or phi 0° of CST) simulations. The measured x-z patterns of linear and planar arrays at different frequency points are shown in Fig. 11.

One of the limitations of the presented CATR system is having only one degree of freedom for the azimuth rotation of the positioner. When measuring antenna arrays, it is required to measure both azimuth and elevation patterns to describe the radiation profile of mmWave antenna array accurately. For instance, for the fan-beam pattern of a linear array, the azimuth plane has a wider HPBW while the elevational plane has a narrower HPBW. To overcome this measurement limitation with a cost-effective solution, and to measure the radiation pattern in the orthogonal plane, we utilized a 2.4 mm standard right angle to rotate the AUT by 90°.

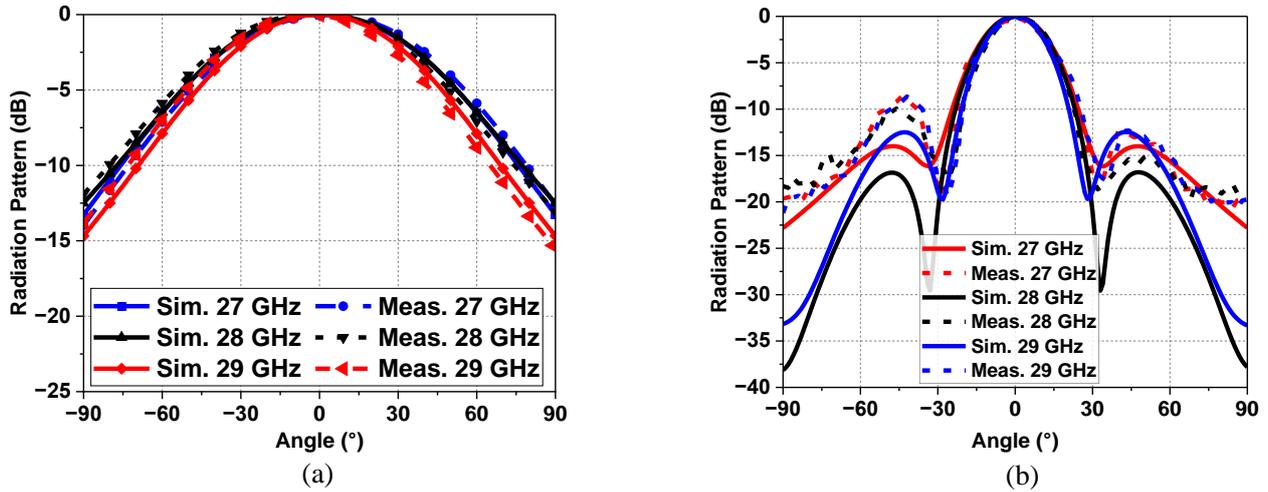

Fig. 11. Measured radiation patterns in x-z plane. (a) Linear array pattern with wider beamwidth. (b) Planar array pattern with narrow beamwidth.

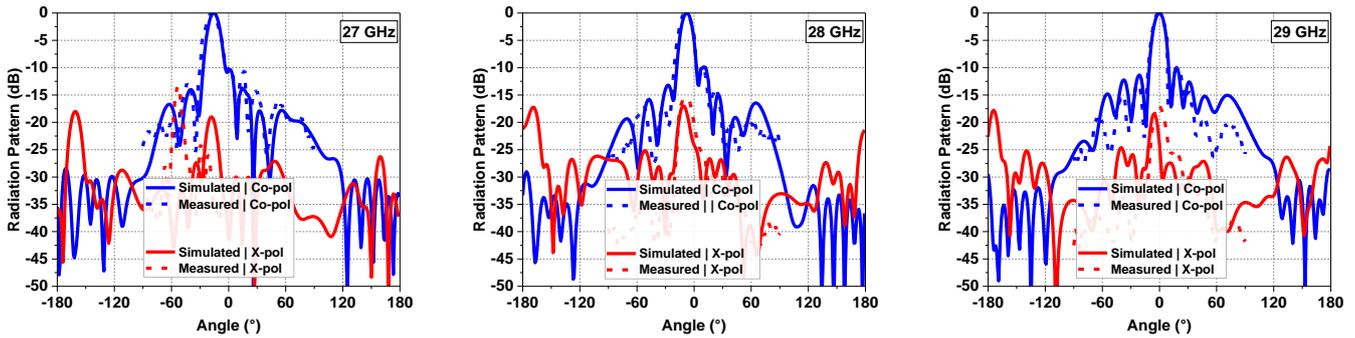

Fig. 12. Measured radiation patterns of the planar array in the y-z plane at different frequency points.

With this approach, the elevational plane pattern (which in this case matches to phi 90° cut of CST simulations) can be measured over a complete 180° by simply rotating the AUT along the horizon, as demonstrated in Fig. 10. However, it is important to mention that while AUT is rotated to 90° (while using a right angle adapter), the Tx cable should also be switched to orthogonal polarization to match the required polarization (co- or x-pol).

The measured radiation patterns of the planar array in the elevation plane are shown in Fig. 12. Both co- and X-pol patterns were measured. The measured HPBW, sidelobe levels, and null depths match the simulated results reasonably well. Minor discrepancies can be further eliminated through reduced angular steps and refined fabrication processes, as surface roughness and fabrication tolerance might have adverse effects, especially at mmWave frequencies. The beam squint is upto -20° due to the inherent characteristics of the series-fed configuration. The variation in measured HPBW lies within 1° whereas the measured null depths and SLLs vary within 3 dB as compared to the simulated results.

Note that here we demonstrated the radiation pattern measurements using two ports of the VNA to provide a pervasive simple solution for measurements (as 2-port VNA is relatively widely available). In the case of 4-port VNA, two of the VNA ports can simultaneously be used for dual-polarized Tx antenna, while the third VNA port can be used for AUT. In that case, S21 and S31 can be captured simultaneously using a customized software program in which one of the transmission coefficients will correspond to co-pol (while the other will correspond to x-pol based, corresponding to the orientation of AUT).

The gain of a mmWave antenna array is an important parameter, as it helps to mitigate path loss and calculate effective isotropic radiated power at the transmitter end. In this work, we measured the realized gain of the AUT by using the gain-transfer (or gain-comparison) method. A mmWave standard gain horn (SGH) antenna with a known gain was used to measure the gain of the AUT. Initially, the relative gain was conducted, and by comparing these measurements to the known gain of the standard antenna, actual gain values of AUT were determined. The process requires two sets of measurements (at each operating frequency point, 0.5 GHz step here) In the first set, the SGH was used as a receiving antenna, and the best maximum received power ($S21_{Horn}$) was recorded. In the second set, the SGH was replaced by AUT, and the received power ($S21_{AUT}$) was recorded. In both sets, the geometric configuration should remain unchanged, except for the replacement of the AUT and SGH, whereas the input power ($P_T$) remains the same (-10 dBm in our measurements).

## Gain Measurement Through Gain-Comparison Method

**Friis Equation**
$$P_R = P_T + G_T + G_R - P_L \quad \text{dB Scale}$$

$P_{AUT}$ = Received power by antenna under test
$P_{Horn}$ = Received power by standard gain Horn antenna
$P_T$ = Transmit power
$G_T$ = Gain of Transmitting Antenna
$G_{AUT}$ = Gain of Antenna Under Test ($G_{AUT}$)
$G_{Horn}$ = Gain of standard gain horn antenna (known gain)
$PL$ = Pathloss

*(Using standard gain horn antenna)*
$$P_{Horn} = P_T + G_T + G_{Horn} - P_L \quad \longrightarrow (1)$$
Note that: $P_{HORN} - P_T = S21_{Horn}$ (dB) on VNA

*(Using AUT)*
$$P_{AUT} = P_T + G_T + G_{AUT} - P_L \quad \longrightarrow (2)$$
Note that: $P_{AUT} - P_T = S21_{AUT}$ (dB) on VNA

Insert value of $G_T$ from equation (1) in equation (2), and solve (2) for $P_{AUT}$

$$P_{AUT} = P_T + (P_{Horn} - P_T - G_{Horn} + P_L) + G_{AUT} - P_L$$
$$(P_{AUT} - P_T) = (P_{Horn} - P_T) - G_{Horn} + \cancel{P_L} + G_{AUT} - \cancel{P_L}$$
$= S21_{AUT}$     $= S21_{Horn}$

$$\boxed{G_{AUT}\,(dBi) = G_{Horn}\,(dBi) + S21_{AUT}(dB) - S21_{Horn}(dB)}$$

Fig. 13. A mathematical depiction of the relative gain measurement method.

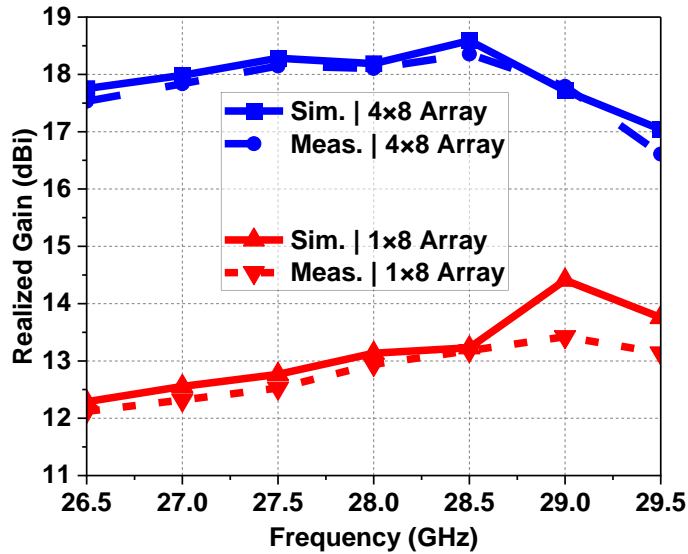

Fig. 14. Measured gain of linear and planar antenna arrays.

Note that in the relative gain measurement method, the effect of path loss (PL) is potentially eliminated, and the system's calibration is not required [37]. An insightful mathematical formulation of the relative-gain method derived from the Friis equation is presented pictorially in Fig. 13. The gain equation reveals that only S21 values (with AUT and SGH) on the VNA are sufficient, provided the gain of the SGH is known. The measured gain matches quite well with the simulated gain as shown in Fig. 14

TABLE 2. Comparison of the proposed antenna performance with some of the other related antennas.

| Ref. | Antenna Topology | -10 dB BW (GHz) | Peak Meas. Gain (dBi) | Peak Radiation Efficiency (%) | Fabrication Complexity |
|---|---|---|---|---|---|
| **[10]** | Single element (4×4 MIMO) | 26.5-30 | 10.5 | 91 | High |
| **[12]** | 1 × 8 array | 24.4-32.2 | 14 | <83 (element) | High |
| **[16]** | Single element (2×2 MIMO) | 26.7-27.9 | 5.63 | NA | Medium |
| [24] | 1 × 8 array | 26.5-29.5 | 13.44 | 70 (average) | Medium |
| [25] | 1 × 8 array | 27-29.5 | 12.85 | NA | Medium |
| [26] | Linear array with inverted cone stubs | 27-28.3 | 10.7 | NA | Low |
| **This work** | 1 × 8 array | 26-30 | 13.4 | 85.77 | Low |
| | 8 × 4 array | 26.5-30 | 18.4 | 85.38 | Low |

## V. CONCLUSION

In this work, we designed a compact, low-cost, and high-gain microstrip antenna array tailored for the 5G n257 FR-2 mmWave band, with a comprehensive demonstration of measurement validation. The 8-element linear array and 32-element array provide a -10 dB impedance bandwidth of 3 GHz around the 28 GHz mmWave band. The peak measured gain of the linear array is 13.4 dBi and that of the planar array is about 18.4 dBi at 28.5 GHz. Both arrays maintain over 83% radiation efficiency. We then elucidated an extensive measurement campaign using the R&S ATS800B® CATR setup to validate the radiation patterns and realized gain of the mmWave array. We demonstrated the effectiveness of the short-range CATR in accurately characterizing antennas operating in the mmWave band, with an excellent match between measured and simulated results. Additionally, we provided a detailed illustration for measuring azimuth and elevation patterns over 180° using a single azimuthal rotator, and we highlighted the procedure for accurately measuring the realized gain of antenna arrays. These findings emphasize the importance of precise antenna testing in confined spaces, particularly in the context of developing high-frequency wireless communication technologies for 5G and beyond.

**Acknowledgment** This work was supported by the CHEDDAR: Communications Hub for Empowering Distributed ClouD Computing Applications and Research, funded by the UK EPSRC under grant numbers EP/Y037421/1 and EP/X040518/1

**Declarations**

**Ethical Approval** Not Applicable
**Data Availability** No datasets were generated during this study
**Competing Interests** The authors declare no competing interests